\documentclass[showpacs,twocolumn,aps,superscriptaddress]{revtex4}
\usepackage{graphicx}
\usepackage{palatino}
\usepackage[latin1]{inputenc}
\usepackage{epsf}
\usepackage{amsmath,amssymb}
\usepackage{latexsym}
\usepackage{calc}
\usepackage{color}
\usepackage{shadow}
\usepackage{epsfig}

\newcommand{\ben}{\begin{equation}}
\newcommand{\een}{\end{equation}}
\newcommand{\bea}{\begin{eqnarray}}
\newcommand{\eea}{\end{eqnarray}}

\def\sss{\scriptscriptstyle\rm}
\def\x{_{\sss X}}
\def\c{_{\sss C}}
\def\s{_{\sss S}}
\def\sup{_{{\sss S},\uparrow}}
\def\sdown{_{{\sss S},\downarrow}}
\def\xcup{_{{\sss XC},\uparrow}}

\def\xup{_{{\sss X},\uparrow}}

\def\cup{_{{\sss C},\uparrow}}

\def\xc{_{\sss XC}}

\def\Hxc{_{\sss HXC}}

\def\H{_{\sss H}}

\def\ext{_{\rm ext}}

\def\ee{_{\rm ee}}

\def\sup{_{{\sss S},\uparrow}}
\def\sdown{_{{\sss S},\downarrow}}
\def\xcup{_{{\sss XC},\uparrow}}

\def\xup{_{{\sss X},\uparrow}}

\def\cup{_{{\sss C},\uparrow}}

\def\br{{\bf r}}

\begin{document}
  \title{Charge-transfer in time-dependent density-functional theory via spin-symmetry-breaking}
  \author{Johanna I. Fuks}
\affiliation{Nano-Bio Spectroscopy group and ETSF Scientific Development Centre, Dpto.~F\'isica de Materiales, Universidad del Pa\'is Vasco, CFM CSIC-UPV/EHU-MPC and DIPC, Av.~Tolosa 72, E-20018 San 
Sebasti\'an, Spain}
\author{Angel Rubio}
\affiliation{Nano-Bio Spectroscopy group and ETSF Scientific Development Centre, 
Dpto.~F\'isica de Materiales, Universidad del Pa\'is Vasco, CFM CSIC-UPV/EHU-MPC and DIPC, Av.~Tolosa 72, E-20018 San 
Sebasti\'an, Spain}
\affiliation{Fritz-Haber-Institut der Max-Planck-Gesellschaft, Faradayweg 4-6,
D-14195 Berlin, Germany}
\author{Neepa T. Maitra}
\affiliation{Department of Physics and Astronomy, Hunter College and the City University of New York, 695 Park Avenue, New York, New York 10065, USA}
  \date{\today}
  \begin{abstract}
Long-range charge-transfer excitations pose a major challenge for
time-dependent density functional approximations.  We show that
spin-symmetry-breaking offers a simple solution for molecules composed of open-shell fragments, yielding
accurate excitations at large separations when 
the acceptor  effectively contains one active electron.
Unrestricted exact-exchange and self-interaction-corrected functionals
are performed on one-dimensional models and the real LiH 
molecule within the pseudopotential approximation to demonstrate our
results.
 \end{abstract}
 \maketitle Although time-dependent density-functional theory (TDDFT)
 has had resounding success in predicting  accurate excitation
 spectra in a wide variety of systems~\cite{RG84,TDDFTbook}, 
 difficulties still plague its application to certain areas.  
The problem of charge-transfer (CT) excitations has drawn especially
significant attention in recent years~\cite{DWH03,T03,GB04bNGB06,
  M05cMT06, TTYY04p,SKB09,HIG09}, due to its relevance for
biomolecules, molecular conductance, solar cell design; these are
systems for which TDDFT would be particularly attractive due to its
favorable system-size scaling. However, it has been challenging to find a satisfactory universal solution to the CT problem:  ab initio approaches based on modeling the exact kernel appear impractical, while practical approaches tend to involve empirical parameters.
 Here we present a new
approach to calculate CT excitations in TDDFT for certain cases, based on spin-symmetry-breaking.
We show that accurate excitations are obtained when the acceptor is an effectively
one-electron system, e.g. an element in Group 1 of the periodic table
treated in the pseudopotential approximation, and justify why this is
so. For large separations, the leading order behavior is captured solely from  Kohn-Sham (KS) orbital energy differences.
Results are given for model systems and for the LiH molecule, and suggest a type of Koopmans' concept for
one-electron systems.

The usual approximations in TDDFT notoriously underestimate CT excitations between fragments at large separation $R$.
To leading order in $1/R$, the exact answer for the lowest
CT frequency is:
\ben
\omega_{\rm CT}^{\rm exact} \to I^D - A^A - 1/R
\label{eq:omexact}
\een 
where $I^D$ is the ionization energy of the donor, $A^A$ is the
electron affinity of the acceptor and $-1/R$ is the first
electrostatic correction between the now charged species.  (Atomic
units are used throughout). It is well-understood why TDDFT severely
underestimates CT~\cite{DWH03,T03,GB04bNGB06,M05cMT06}: In TDDFT, the
first step is to compute the Kohn-Sham (KS) orbital energy differences
between occupied ($i$) and unoccupied ($a$) orbitals, $\omega\s = \epsilon_a - \epsilon_i$. In a second
step, these frequencies are corrected to the true excitations via the
Hartree-exchange-correlation kernel, $f\Hxc[n_0](\br,\br',\omega)$, which shifts and mixes the KS excitations within a matrix formulation. The kernel is 
a
functional of the ground-state density $n_0(\br)$, with matrix
elements $\int d^3r d^3r'
\phi_i(\br)\phi_a(\br)f\Hxc(\br,\br',\omega)\phi_{i'}(\br')\phi_{a'}(\br')$. 
For
CT excitations, the vanishing spatial overlap at large separations between
occupied donor and unoccupied acceptor orbitals sitting on different
nuclei means that the TDDFT predictions for CT excitations reduce to the KS orbital
energy difference, $\epsilon_a - \epsilon_i$ when using usual semi-local
functional approximations for $f\Hxc$.
With approximate ground-state functionals the
highest occupied molecular orbital (HOMO) energy, $\epsilon_{\rm H}$,
underestimates the true ionization energy, while the lowest unoccupied
molecular orbital (LUMO), $\epsilon_{\rm L}$, lacks relaxation
contributions to the electron affinity.  
%
The last few years have seen many  methods to correct the
underestimation of CT excitations, e.g. Refs.~\cite{TTYY04p,SKB09,HIG09}; most
modify the ground-state functional to correct the approximate KS
HOMO's underestimation of $I$, and mix in some
degree of Hartree-Fock, and most, but not all~\cite{SKB09,HIG09} determine this mixing via at least one empirical parameter. 
 Fundamentally, staying within pure DFT, both the relaxation contributions to $A$ and the
$-1/R$ tail in Eq.~\ref{eq:omexact} come from $f\Hxc$, which must
exponentially diverge with fragment
separation~\cite{GB04bNGB06,M05cMT06}.  Worse, in the case of
open-shell fragments, {\it not} covered by most of the recent fixes,
additionally the exact $f\xc$ is strongly
frequency-dependent~\cite{M05cMT06}.

The major reason for the awkward kernel structure in the case of
open-shell fragments lies in the KS ground-state description: the HOMO
(and LUMO) are delocalized over the whole molecule, quite distinct
from the Heitler-London-like nature of the true wavefunction. In either the case of the exact or semi-local approximations, 
their orbital energy difference tends to zero as the molecule is pulled apart, and so $f\Hxc$ must be responsible for the entire CT energy~\cite{M05cMT06}.
 It is
long-recognized that this static correlation error~\cite{MCY09} is the
root of the problem of poor {\it ground-state} energies, studied
extensively in molecules like H$_2$~\cite{GL76p}, and that a
simple way out  is to allow the system to break
spin-symmetry.  An unrestricted calculation with an approximate
functional run on a diatomic molecule, leads, at a critical internuclear
separation, to the spin-polarized solution
obtaining the lowest ground-state energy.  Albeit having incorrect
spin-symmetry, accurate ground-state energies are achieved essentially
because the KS description is rendered to have one electron on each
atom.

Although less discussed, the same physics applies for heteroatomic
molecules composed of open-shell fragments~\cite{TMM09}, and suggests
that symmetry-breaking could be a means to obtain its CT excitations.
If the exact functional were
used, the lowest-energy state remains correctly spin-unpolarized at
any $R$, but at the cost of stark step and peak features  in
the bonding region, and strong-frequency-dependent structure in
$f\Hxc$, difficult for approximations to capture. If instead, correct
spin-symmetry is imposed on any existing approximate density-functional, the
ground-state displays unphysical fractional charges at large $R$,
delocalized HOMO and LUMO orbitals, and again poor CT
energies~\cite{M05cMT06,TMM09}.

The following examples  show that
remarkably accurate CT excitations can indeed be obtained from TDDFT
via spin-symmetry breaking when the acceptor contains effectively one active electron; for large enough separations, these are
contained in simply the bare KS excitations.  
We present first one-dimensional models that enable us
to compare with highly accurate numerically-exact solutions (computed
using a Runge-Kutta differential equation solver as implemented in the
octopus code~\cite{octopus}), and to then analyze and understand in
detail why the CT in symmetry-broken TDDFT is accurate via the
underlying potentials. For the DFT calculations we study the
exact-exchange (EXX) and local spin density
approximation(LSD)~\cite{CSS06} with self-interaction correction
(SIC).
 We shall see their correct long-range behavior yields the
correct $R$-dependence at large separations. We stress that this long-rangedness must be combined with symmetry-breaking in order to get accurate CT excitations: restricted calculations using these functionals fail~\cite{CGGG00}. The functionals are implemented in
octopus within the KLI approximation to OEP~\cite{KLI}.  After
studying the models, we then turn to the real LiH molecule.
In our first model, the nuclear potentials are represented by short-range wells, at separation $R$:
\ben 
v\ext(x) = 
-U^L/\cosh^2(x) - U^R/\cosh^2(x-R) \;,
\label{eq:Vextcoshsq}
\een
where the strengths $U^L = 5.5$ au, and $U^R=6$au. We place
two electrons interacting soft-Coulombically via
\ben
v\ee(x,x') =1/\sqrt{1+(x-x')^2}\;.
\label{eq:softCoulomb}
\een
into this heteroatomic molecule.
The unrestricted KS calculation yields a
spin-unpolarized solution until a separation of about $R\sim 2.3$a.u.
(coinciding with an avoided crossing in the potential energy
surfaces~\cite{TMM09}) where it begins to break symmetry: the
eigenvalues for the up and down spin in the ground-state begin to
separate and slowly approach those of the isolated wells as the
molecule is pulled further apart and each electron settles in a
different well.  In contrast to  Ref.~\cite{KMK08},
localization of orbitals is in fact achieved within KLI: for two electrons
KLI is equivalent to full OEP, but even for systems of more than two electrons (not shown here) KLI yielded spin-polarized localized orbitals. 

\begin{figure}[h]
\centering
\includegraphics[height=4.2cm,width=8.5cm]{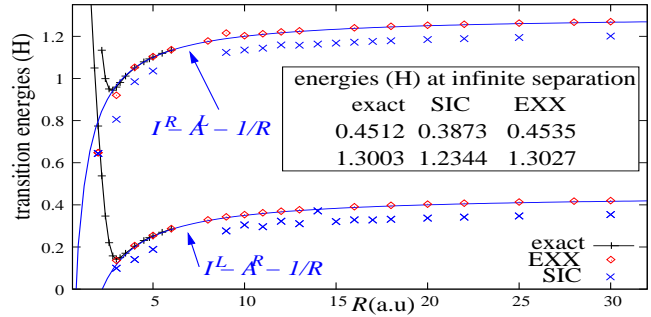}
\caption{Charge-transfer excitation energies for the model Eq.~(\ref{eq:Vextcoshsq})-(\ref{eq:softCoulomb}) using unrestricted SIC and EXX. The exact converge onto the asymptotic Eq.~\ref{eq:omexact} (solid blue lines) soon after the symmetry-breaking point. (Color online)}
\label{fig:CT_softCoul}
\end{figure}

Figure~\ref{fig:CT_softCoul} plots the lowest orbital excitation
energies of the model
Eq.~(\ref{eq:Vextcoshsq})-(\ref{eq:softCoulomb}).
 The KS energy differences, especially those of EXX, capture the
exact CT excitations throughout with remarkable accuracy! 

\begin{figure}[h]
\centering
\includegraphics[height=4cm,width=8cm]{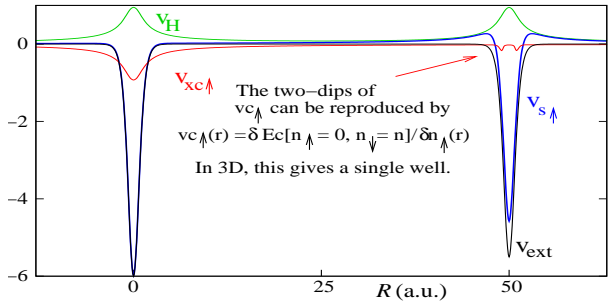}
\caption{The KS potential $v\sup$ (blue) and its components. (Color online).}
\label{fig:vsup}
\end{figure}

Why this is so can be seen by studying the underlying KS
potentials. Consider first the limit $R\to\infty$, where symmetry-breaking has placed, say,  the spin-up(down)
electron in the left(right) well in the ground state. The left well is the donor for the following discussion.
Fig.~\ref{fig:vsup} plots the KS potential for the $\uparrow$-spin:
\ben
v\sup[n_\uparrow,n_\downarrow] = v^A\ext + v^D\ext+ v\H[n_\uparrow]+ v\H[n_\downarrow] + v\xcup[n_\uparrow,n_\downarrow] 
\label{eq:vsup}
\een
and its components at a separation of
$R=50$a.u. In Eq.~\ref{eq:vsup}, $v^{A(D)}\ext$ refers to the atomic acceptor(donor) potential (i.e. the right(left) well), 
and $v\H[n](\br) = \int d^3r' n(\br')/|\br - \br'|$ is the Hartree potential generated by density $n(\br)$.
As $R\to\infty$, in the vicinity of the donor $v\cup = 0$ and $v\xup = -v\H[n_\uparrow]$, so $v\sup = v^D\ext$: the system is
essentially a one-electron system in this limit and local excitations of the $\uparrow$-electron are, correctly, just excited states of the donor. 
In the vicinity of the acceptor, where only a $\downarrow$-electron
lives, $v\xup \to 0$ in the limit $R\to\infty$, and
\ben
v\sup^\infty(\br \sim {\rm acceptor}) = v^A\ext + v\H[n_\downarrow] +v\cup[n_\uparrow,n_\downarrow] \;.
\label{eq:vsupother}
\een
The Hartree potential generated by the $\downarrow$-electron and a
small correlation contribution (the two dips in the red curve)  result in a net upward
shift of $v\ext$: the unoccupied $\uparrow$-electron states living in
the right-hand-well are shifted up in energy compared with those of
the $\downarrow$-electron (i.e. those of $v\ext$). An approximate {\it
  affinity} level is thereby induced in the right-hand-well. So the bare KS orbital energy-differences yield accurate CT
excitations: first, the HOMO for the $\uparrow$-electron is the lowest orbital in the left-hand-well, for which $\epsilon_{\rm
  H} = -I^D$, the exact ionization potential, due to Koopmans' theorem~\cite{Koopmans}, since both EXX and SIC are
exact for one electron.
Second, and more significantly, the LUMO approximates the
affinity level of the right-hand-well, $\epsilon_{\rm L} \approx
-A^A$, sensing the presence of the $\downarrow$-electron, i.e. Hartree-correlation relaxation contributions to the affinity are already incorporated at the bare KS level.
We call
this eigenstate of Eq.~\ref{eq:vsupother}, and the corresponding state
of $v\sdown$, for which entirely analogous analysis holds, the
``induced affinity'' levels of the right and left atoms respectively. In the limit of infinite separation, they
converge onto the lowest state for the unoccupied spin of the isolated
one-electron atom.  A key point is that $v\sup$ and $v\sdown$ are
different: in any restricted calculation where these are the same,
unoccupied levels are {\it excitations} (of the same $N$-electron
system), not {\it affinities}.


How well does the induced affinity level approximate the true
affinity generally? DFT theory tells us that the exact affinity of an
$N$-electron atom is
$A(N) = E(N) - E(N+1) = -\epsilon_{\rm H}(N+1)$
where the middle expression is computed from total energy
differences while the third expression  is computed from the highest occupied KS eigenvalue of the relaxed $(N+1)$-electron system~(see eg.~\cite{GKKG00p}). 
Let us then consider the KS potential of a two-electron atom. Because the density is localized, an unrestricted calculation yields a spin-unpolarized result. Denoting the two-electron ground-state density $n_2(\br)$, 
\ben
v\s[n_2]= v\ext + v\H[n_2/2]+ v\c[n_2] 
\label{eq:vstwoe}
\een 
where $v\ext$ is the nuclear potential and we have noted that, for exact exchange,
$v\x[n_2] = -v\H[n_2]/2 = -v\H[n_2/2]$.  
First, neglecting
correlation, Eq.~\ref{eq:vstwoe} is very close to
Eq.~\ref{eq:vsupother} {\it if} $n_2 \approx 2n_\downarrow$, i.e. if, when a
second electron is added to a well in which there is already one
electron, there is little density relaxation. This is the case in the
 model example above, since the dominant part of the energy of
the electrons is from the external potential.
In such a case, Eqs.~\ref{eq:vsupother} and \ref{eq:vstwoe}
then imply that the induced affinity level approximates the true
affinity well.   It will be a lower bound (i.e. $\vert
\epsilon_{L,\uparrow}\vert \le \vert \epsilon_H(N=2)\vert$), because
electron repulsion leads to $v\H[n_2/2]$ being a little weaker than
$v\H[n_\downarrow]$. This was borne out in all the EXX results of
different models we considered. Correlation tends to raise the induced affinity, sometimes bringing it higher than the true affinity: certainly using LSD-SIC, $v\cup[n_\uparrow=0,n_\downarrow =n]$ forms a deeper negative well than $v\c[n_2]$. Shortly we will discuss examples in which the density relaxation is important, so that $n_2$ is not very close to $2n_\downarrow$, and there the affinity level is not such a good approximation to the true affinity; consequently the charge-transfer excitations are not as accurate. 
 
The arguments above hold only for one-electron acceptors: if the
acceptor already has an electron of the transferring spin in it, then
excitations of that spin are the usual constant-number excitations of
TDDFT, not approximate affinity levels. The donor however may contain any
number of electrons: similar models that have, for example, three electrons in one well and one in another again showed excellent CT excitations from the former to the latter, under spin-symmetry-breaking.

We now extend the discussion following Eq.~\ref{eq:vsup} to the case
of an $N$-electron donor and 1-electron acceptor, at finite but large
separation. For ease of notation, assume again the acceptor carries a
$\downarrow$-electron in the molecular ground state.  First consider
Eq.~\ref{eq:vsup} near the $N$-electron donor.  For external (nuclear)
potentials that decay Coulombically, the ($N$-electron), $v\sup \sim
v\s^D + O(1/R^3)$, where $v\s^D$ is the KS potential of the donor
atom. At the acceptor, $v\sup \sim v\sup^\infty -1/R +
O(1/R^3)$ (noting that $v\ext^D$ cancels $v\H[n_\uparrow]\approx N^D/R$, while
$v\xup \sim -1/R$). So, to leading order in $R$, for the $\uparrow$-electron, $\epsilon_{\rm H} = \epsilon_{\rm H}^D$ and $\epsilon_{\rm L} = \epsilon_{\rm L}^\infty - 1/R$ where $\epsilon_{\rm L}^\infty$ is the induced affinity level of the acceptor in the infinite separation limit. 
Therefore, 
\ben
\omega\s = \epsilon_{\rm L} - \epsilon_{\rm H} = \epsilon_{\rm L}^\infty - \epsilon_{\rm H}^D - 1/R\;,
\een
as in Eq.~\ref{eq:omexact}, with 
$I^D\approx -\epsilon_{\rm H}^D$ and $A^A$ approximated by the induced affinity level. 
(For short-ranged potentials as in Eq.~\ref{eq:Vextcoshsq}, the arguments lead instead to $\epsilon_{\rm H} = \epsilon_{\rm H}^D+1/R$ and $\epsilon_{\rm L} = \epsilon_{\rm L}^\infty$). Therefore, a long-ranged exchange-correlation as in EXX or SIC, once symmetry-broken, yields good CT excitations, from just its bare KS orbital energies, as a function of $R$, for large $R$. Note the importance of correct asymptotics of the functional used for correct $R$-dependence, as well as for accurate ionization potentials and induced affinities. 

\begin{figure}[h]
\centering
\includegraphics[height=4cm,width=7cm]{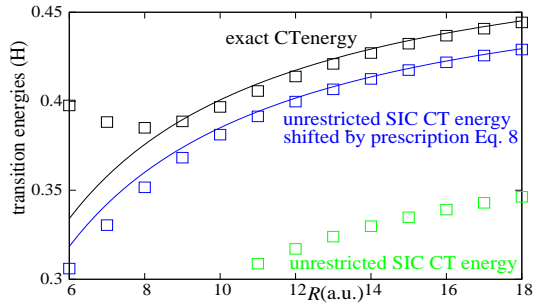}
\caption{CT from the left to the right well of model Eqs.~(\ref{eq:softCoulomb})-~(\ref{eq:1dSoftCoul}). Exact: $I^D = 0.6206$H and $A^A = 0.1199$H. Unrestricted SIC: $I^D = \epsilon^D_H = 0.6206$H, $A^A = 0.1355$H, and $\epsilon_L^{\infty,A} = 0.2183$H. (Color online)}
\label{fig:softCoulCT}
\end{figure}

For an accurate CT asymptote the density
relaxation upon the addition of an electron must be small. But even when
density relaxation effects are significant, symmetry-breaking can
still be useful as we now explain.
In a practical sense, the infinite-separation limit itself is not so
much a problem for TDDFT, because total ground-state energy differences computed
from DFT~\cite{LFB10} can often yield reliable values for $I$ and $A$
in Eq.~\ref{eq:omexact}. Rather it is intermediate but large distances that are the challenge, where CT energies deviate from the asymptotic formula Eq.~\ref{eq:omexact}. Our symmetry-breaking approach can capture these deviations, going beyond Eq.~\ref{eq:omexact}. The procedure is to compute the (symmetry-broken) KS HOMO and LUMO energy difference, but, when density-relaxation is large, shift it by
\ben
I^D - A^A - (\epsilon_{\rm L}^{\infty, A} - \epsilon_{\rm H}^D) 
\label{eq:shift}
\een
where $I$ and $A$ are computed from total ground-state DFT energy differences. 
In this way, asymptotically, the curves approach
Eq.~\ref{eq:omexact} accurately, but for intermediate to large distances, contain
correct physical deviations from Eq.~\ref{eq:omexact} due to
polarization. 
To illustrate this, we now consider soft-Coulomb nuclear wells: 
\ben
V\ext = - 1/\sqrt{x^2 + a_L}-  1/\sqrt{(x-R)^2 + a_R}
\label{eq:1dSoftCoul}
\een
and Fig.~\ref{fig:softCoulCT} takes $a_R=0.7$ and $a_L=1.2$.
The more diffuse densities of these wells makes them more polarizable so deviations from Eq.~\ref{eq:omexact} are more evident, as shown in  Fig.~\ref{fig:softCoulCT}: at intermediate separations, the (exact) CT energies fall shy of the asymptotic Eq.~\ref{eq:omexact}, shown as the black curve, due to the local polarization of the CT state towards the positive charge at the other nucleus.
After applying the shift of Eq.~\ref{eq:shift} the unrestricted SIC results approach the exact results well, and capture this attractive shift; this holds also for CT in the other direction (not shown). (The blue curve is the asymptote for the unrestricted SIC prediction, but also shifted according to Eq.~\ref{eq:shift}).  

So far, we have discussed CT excitations obtained from bare KS
excitations alone and argued why these work so well, as demonstrated
by the model examples. The second step of TDDFT is to apply $f\Hxc$ to
correct the KS excitations towards the exact ones; there are both ``diagonal'' terms which shift each KS excitation, as well as ``off-diagonal'' ones that mix them.  
 For the systems so
far discussed we expect both these effects are small, because (i) the diagonal term involves overlap of the occupied and unoccupied orbitals in the excitation, which vanish exponentially at large distances,  and (ii) there is little mixing
with other excitations in the system. Mixing and shifting of KS excitations will be
important at small and intermediate distances,  
leading to further deviations from the
asymptotic Eq.~\ref{eq:omexact}, especially for real molecules, given their higher density of states.

\begin{figure}[h]
\centering
\includegraphics[height=5cm,width=8cm]{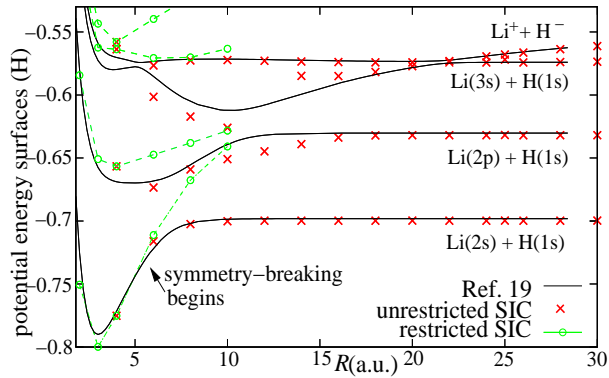}
\caption{Potential energy surfaces of the LiH molecule. Even the (shifted) bare KS energy-differences of unrestricted SIC are close to the CI ones of Ref.~\cite{ADD09}. (Color online)}
\label{fig:LiH}
\end{figure}

Turning now to a real molecule, LiH: Using a pseudopotential for the
Li atom renders it an effectively one-electron atom, and our method
captures CT in both directions.  In Fig.~\ref{fig:LiH} we plot the
lowest potential energy surfaces of the LiH molecule, computed with
unrestricted SIC, with the Troullier Martins pseudopotential coded in octopus~\cite{octopus}, compared to the highly-accurate
configuration-interaction (CI) calculations of Ref.~\cite{ADD09}.  The
induced affinity level of H is 0.0726H while that computed from
ground-state DFT energy-differences is 0.0264H, closer to the experimental
(0.0277H), so we have applied the shift of Eq.~\ref{eq:shift} for
$R\ge 12$. As accurate energies are unavailable for CT from H to Li, we
do not show this curve here. The excellent agreement of the
unrestricted approach with CI  can be
contrasted with restricted SIC calculations, whose collapse at smaller
$R$ is due to the near-degeneracy of the HOMO and LUMO mentioned earlier; the latter leads to convergence difficulties for larger separations. 
The symmetry-broken SIC predicts the separation at which there is a crossing between the ionic curve and the Li(3s) curve very accurately, although it
appears more as a direct crossing rather than an avoided one. 

In summary, we have shown that symmetry-breaking is a simple
non-empirical way to obtain CT excitations from KS orbital energies alone, for acceptors
that contain effectively one electron, and explained why.  Strikingly
good results were obtained for model systems as well as for real
molecule LiH and further studies are underway.  Applying the TDDFT kernel should improve the accuracy at intermediate distances, capturing mixing of CT and local excitations, while these $f\Hxc$ corrections will vanish asymptotically.

Conceptually, the observation that
for one-electron systems, the levels of the unoccupied spin
approximate affinity levels can be interpreted in an
extended Koopmans' sense. Koopman's theorem in DFT states that
$I=-\epsilon_{\rm H}$ exactly, while generalized Koopman's theorem
applies to Hartree-Fock where $I \approx -\epsilon^{\rm HF}_{\rm H}$
and $A \approx -\epsilon^{\rm HF}_{\rm L}$, leading to the use of
hybrid functionals for CT, mentioned earlier~\cite{TTYY04p,SKB09}. Although the
LUMO in exact DFT represents an excitation of the $N$-electron system, rather than the $(N+1)$-electron one, our results show that when $N=1$ in
spin-DFT,  the levels of the unoccupied spin can be interpreted in a generalized Koopmans' sense, as they  approximate
affinity levels.

We acknowledge support from MEC (FIS2007-65702-C02-01), ACI-promociona  (ACI2009-1036),
Grupos Consolidados UPV/EHU del Gobierno Vasco (IT-319-07), e-I3 ETSF project (Contract No. 211956), the National Science Foundation (CHE-0647913), the Cottrell Scholar Program of Research Corporation and a grant of computer time from the CUNY High Performance Computing Center under NSF Grants CNS-0855217 and CNS-0958379.

\end{document}